\documentstyle[12pt,aaspp4]{article}
\input psfig.sty

\begin{document}

\title{The Frequency of Active and Quiescent Galaxies with Companions:
Implications for the Feeding of the Nucleus}

\author{Henrique R. Schmitt\altaffilmark{1,2,3}}
\altaffiltext{1}{National Radio Astronomy Observatory, P.O. Box 0,
1003 Lopezville Road, Socorro, NM\,87801}
\altaffiltext{2}{Jansky Fellow}
\altaffiltext{3}{email: hschmitt@aoc.nrao.edu}

\date{\today}

\begin{abstract}

We analyze the idea that nuclear activity, either AGN or star formation,
can be triggered by interactions, studying the percentage of active,
HII and quiescent galaxies with companions.
Our sample was selected from the Palomar survey, and avoids selection biases
faced by previous studies. This sample was split into 5 different groups,
Seyfert galaxies, LINERs, transition galaxies, HII galaxies and absorption
line galaxies. The comparison between the local galaxy density distributions
of the different groups showed that in most cases there is no statistically
significant difference among galaxies of different activity types, with the
exception that absorption line galaxies are seen in higher density
environments, since most of them are in the Virgo cluster. The comparison of
the percentage of galaxies with nearby companions showed that there is a
higher percentage of LINERs, transition, and absorption line galaxies with
companions than Seyferts and HII galaxies. However, we find that when we
consider only galaxies of similar morphological types (ellipticals or spirals),
there is no difference in the percentage of galaxies with companions among
different activity types, indicating that the former result was due to the
morphology-density effect. Also, only small differences are found when we
consider galaxies with similar H$\alpha$ luminosities. The comparison
between HII galaxies of different H$\alpha$ luminosities shows that there is
a significantly higher percentage of galaxies with companions among HII
galaxies with L(H$\alpha$)$>10^{39}$~erg~s$^{-1}$, than among those with
L(H$\alpha$)$\leq10^{39}$~erg~s$^{-1}$, indicating that interactions increase
the amount of circumnuclear star formation, in agreement with previous results.
The fact that we find that galaxies of different activity types have the
same percentage of companions, suggests that interactions between galaxies
is not a necessary condition to trigger the nuclear activity in AGNs.
We compare our results with previous ones and discuss their implications.

\end{abstract}


\section {Introduction}

Two of the major concerns in the study of AGNs are the origin of the gas
which fuels the nuclear black hole and the mechanisms responsible for making
the gas lose angular momentum and move from galactic scales down to the
inner $\sim$1 pc region of the galaxy. In spiral galaxies, gas in their
disks can be a natural source to fuel the AGN. Several mechanisms have
been suggested to explain how it is possible to transport gas from the
disk of a spiral galaxy to its nucleus, like interactions (Gunn 1979;
Noguchi 1987,1988; Hernquist 1989; Barnes \& Hernquist 1992), or
bars (Schwartz 1981; Norman 1987; Shlosman, Frank \& Begelman 1990).
In the case of elliptical galaxies the situation is more complicated,
because they only have small quantities of gas, which implies that the gas
which feeds their nuclei most likely has an origin external from the galaxy.
Reviews about this subject, with detailed explanations of the possible
mechanisms responsible for feeding the nucleus are given by Shlosman et al.
(1990) and Combes (2001).

The influence of interactions on the fueling of the nucleus of AGNs as well
as HII galaxies has been the topic of several studies. From the theoretical
point of view and N-body simulations, Noguchi (1987,1988), Hernquist (1989),
Barnes \& Hernquist (1992), Byrd et al. (1986), Byrd, Sundelius \& Valtonen
(1987), Lin, Pringle \& Rees (1988), Hernquist \& Mihos (1995),
Mihos \& Hernquist (1994), Taniguchi \& Wada (1996), among others,
have shown that interactions between galaxies, mergers, and minor mergers can
be responsible for bringing gas from the disk to the nuclear region, where it
is shocked and compressed, producing a starburst and subsequently feeding
the nucleus. This idea is qualitatively confirmed with observations, but the
models still have to explain fine details.
It is a well established fact that luminous galaxies, in particular luminous
and ultraluminous infrared galaxies are closely related
to interacting systems (Sanders et al. 1988; Lawrence et al. 1989;
Leech et al. 1994; Gallimore \& Keel 1993). It has also been shown by
Keel et al. (1985), Wright et al. (1988) and Kennicutt et al. (1987)
that interacting galaxies have a higher level of H$\alpha$, infrared and
radio emission, as well as an increased nuclear star formation rate. 

However, in the case of lower luminosity sources, like Seyfert galaxies,
the situation is not so clear. There is no consensus on how important
is the effect of interactions in the feeding of their nuclei, and if
these galaxies have an excess of companions. Stauffer (1982) was one
of the first to point out that Seyferts usually occur in groups.
Dahari (1984, 1985) made a large compilation of possible companion
galaxies around Seyferts. He found that these galaxies have an excess of
companions relative to normal galaxies. Although this result have been
confirmed by several other studies (Rafanelli, Violato \& Baruffolo 1995;
Laurikainen et al. 1994; MacKenty 1989, 1990), there have been some
claims to the contrary. Fuentes-Williams \& Stocke (1988), Bushouse
(1986, 1987), Ulvestad \& Wilson (1984, 1985)
and De Robertis, Yee \& Hayhoe (1998) did not find a detectable 
excess of companions around Seyferts. Other studies of Seyfert galaxies have
shown yet another result, where Seyfert 2's have a larger number of companions
when compared to normal galaxies, while Seyfert 1's do not (Laurikainen
\& Salo 1995 and Dultzin-Hacyan et al. 1999). Similarly, in a study of
galaxies with activity types other than Seyfert, Pastoriza et al. (1994)
showed that galaxies in high and low density media have similar stellar
populations, while Telles \& Maddox (2000) found that low redshift HII
galaxies are, on average, found in slightly lower density environments than
normal field galaxies. Taylor et al. (1995) found a different result, but
according to Telles \& Maddox (2000) this can be due to the fact that
the progenitors of HII galaxies are different from those of field galaxies.

The different results obtained by different papers are most likely due
to the way they selected their samples and control samples, as
pointed out by Heckman (1990) and Osterbrock (1993). In order to do a proper
comparison between the percentage of active, normal and HII galaxies with
companions, it is necessary to have a sample of active galaxies and a control
sample with matching host galaxy properties, which may have been a problem in
the earlier studies. Another problem with these studies is that they usually
did not have information about the redshifts of the companion galaxies.
In most of the cases, the only information available was the magnitudes of
the possible companions and their projected distances to the galaxies being
studied. This means that these companions may not necessarily be physically
associated. The way around this problem was to apply statistical corrections
for the number of background and foreground objects around the galaxies
being studied.

Here we try to solve these problems by studying the percentage of
companion galaxies in the Palomar survey (Ho, Filippenko \& Sargent 1997a).
We will address problems related
to the selection of the sample, control sample, and the availability of 
redshifts for the companions galaxies. We will also be able to study the
percentage of companion galaxies in galaxies of different activity types,
not only Seyferts and normal galaxies as done in previous papers.

This paper is organized in the following way. Section 2 presents
the sample and corresponding subsamples, while Section 3 gives the
techniques and data used. Section 4 presents the comparison
of local galaxy densities for galaxies of different activity types,
and Section 5 shows the comparison of the percentage of companions
in these galaxies as a function of different properties (e.g. morphological
type, H$\alpha$ luminosity). Section 6 presents the discussion of the
results and the summary.

\section{The sample}

As pointed out in the previous section, one of the main problems for the
previous works in this subject was the selection of the active and
corresponding control samples of galaxies. Here we solve this problem
by using the galaxies in the Palomar survey (Ho et al. 1997a).
This sample is ideal for this kind of study, because it contains
the galaxies brighter than B$_T=12.5$ mag in the northern hemisphere.
The relative advantages of this sample relative to others is discussed by
Ho \& Ulvestad (2001). They also point out selection biases which affect
other samples commonly used in the literature. Here we summarize the 
advantages of the Palomar survey. The most important characteristic of this
sample is the fact that it provides homogeneous high-quality spectroscopic
measurements of emission line fluxes, as well as activity classification of
the galaxies. Another of its advantages is the fact that it covers
a large range of AGN and HII luminosities, as well as a large
range of host galaxy parameters, like morphological types and absolute
magnitudes, besides the fact of being large enough to ensure a significant
comparison between galaxies of different activity types as a function of
these parameters. Since we are going to compare the percentage
of galaxies with companions among different activity types, not only
Seyferts and normal galaxies, this sample is ideal for this work, because
it includes both the AGN and control sample in itself, allowing us to
determine the frequency of galaxies with companions in the local universe,
avoiding the selection effects faced by previous works.

The initial sample studied by Ho et al. (1997a) contained 486 galaxies.
However, we decided to exclude local group galaxies according the list
from Mateo (1998), because it includes a large number of low luminosity
dwarf spheroidals.
Galaxies with B$_T>12.5$ mag, which were included in their survey for
historical reasons, or because they were interesting objects, have also been
excluded. This gives a total number of 451 galaxies, divided
into 46 Seyferts, 193 HII galaxies, 88 LINERs, 63 transition objects
and 61 absorption line galaxies. Transition objects are galaxies which
present emission line ratios intermediate between those of HII galaxies
and LINERs, and absorption line galaxies are those without any detectable
emission lines. Other galaxies in this sample had emission line ratios which
resulted in different classifications, depending on the emission line
ratios used. They were classified by Ho et al. (1997a) as L2/S2, or S2/H
for example, and in these cases we used the first of the two classifications.
Given the small number of Seyferts and LINERs with broad emission
lines in the sample, we do not attempt to analyze these galaxies separately.
Nevertheless, we point out that Schmitt et al. (2001) have shown, using
a well defined sample of Seyfert galaxies, that there is no significant
difference in the percentage of Seyfert 1's and Seyfert 2's with companion
galaxies.

A statistical comparison of the emission line and host galaxy properties of
these galaxies as a function of activity type was presented by Ho, Filippenko
\& Sargent (1997b). It should be noticed that most of the galaxies classified
as HII nuclei present only low levels of current star formation, with
H$\alpha$ luminosities similar to those of individual giant HII regions.
Due to this fact, the lower luminosity HII galaxies should be considered as
quiescent galaxies and not as starburst. 

\section{Techniques and data}

We used two different techniques to compare the environments of the different
types of galaxies. The first one used the local galaxy densities defined by
Tully (1988), $\rho_{gal}$, and compared the galaxies in the different
activity classes using the Kolmogorov-Smirnov test (KS-test). $\rho_{gal}$
is the density of galaxies brighter than M$_B=-16$ mag in the vicinity
of the object of interest. This value was calculated by Tully (1988), using
all galaxies in the local universe with distances smaller than 40~Mpc,
taking into account completeness effects. The values used in this paper
were obtained from Table 10 of Ho et al. (1997a), and cover 92\% of the
sample.

The second technique searched for companions around the galaxies using NED
and the Digitized Sky Survey images (DSS).
Several parameters have been used to determine if a
galaxy has a companion or not. We adopt here a slightly modified version of
the one used by Rafanelli et al. (1995), which was determined based
on catalogs of interacting and pair galaxies. We consider a galaxy as a
companion if its distance to the galaxy of interest is smaller than 5 times the
diameter (D$_{25}$) of that galaxy, the difference in brightness between them
is smaller than 3 magnitudes ($|\Delta m|\leq$ 3 mag), and the difference in
radial velocities is smaller than 1000 km~s$^{-1}$ ($ | \Delta \rm Vel | \leq$
1000 km s$^{-1}$). We will call the
galaxy around which we search for companions the primary and the companion
galaxy secondary. The original parameters used by Rafanelli et al. (1995)
considered only secondaries at a distance smaller than 3 times the diameter
of the primary. We decided to increase this value
to 5 times the diameter of the primary. We will show below that this
different criterion does not make a significant difference in the results.

We emphasize that these criteria used to search for companions around
the galaxies are simply empirical, corresponding to the range of properties
found in galaxy pairs. According to Charlton \& Salpeter (1991), pairs with
velocity differences smaller than 150 km~s$^{-1}$ are likely to be bound.
However, the galaxies do not need to be bound in order to interact, the
interaction can happen between galaxies which are not gravitationally bound.

Based on the magnitude criterion alone, we can calculate that, for an
M$^*$ primary galaxy, with M$_B^*=-20.3$ (Binney \& Merrifield 1998)
a secondary galaxy 3 mag fainter will have M$_B=-17.3$, a value between
that of the LMC and the SMC, which have M$_B=-18.5$ mag and M$_B=-16.3$ mag,
respectively. The LMC and SMC magnitudes were calculated using the values
from Mateo (1992) and the distance moduli obtained by Cioni et al. (1992).
This shows that the search for companions will not be biased towards bright
galaxies only. The most luminous galaxy in the sample is NGC\,4594, which
has M$_B^0=-23.12$, and the search for companions around this galaxy will
consider only the galaxies with absolute magnitudes of the order of
M$_B^*$. On the other hand, the faintest galaxy in the sample is NGC\,404,
which has M$_B^0=-15.98$ and companions to this galaxy could have luminosities
as faint as M$_B^0\sim-13$ which is a value typical of Dwarf Spheroidal
galaxies like Fornax and Leo (Binney \& Merrifield 1998).

Most of the galaxies in the sample have M$_B<-19$ (Ho et al. 1997b),
which means that our criteria will select only secondaries with absolute
magnitudes similar to that of the SMC. Lower luminosity
Dwarf/Spheroidal and Irregular galaxies, like the ones on the Local Group,
will be missed.  It was suggested by De Robertis et al. (1998) that mergers
of gas-rich disk galaxies with satellite companions (dwarf galaxies) may
be an important way to trigger the activity in Seyfert galaxies.
The data and criteria used in this paper does not allow us to test
this hypothesis. It may be possible to do such a search in the future,
using results from the SLOAN survey.

One of the problems with previous studies was the fact that they
were usually not able to obtain redshifts of the supposed companion galaxies.
Due to this fact they had to rely on statistical methods to determine the
probability of two galaxies being companions, as well as to correct for the
number of background and foreground galaxies based on galaxy number counts.
Since we cut our sample at B$_T=12.5$ mag, the faintest secondary galaxies
must have magnitudes B$_T=15.5$ mag. In this way we are
sure that most of the galaxies will have redshift information.
Redshift surveys, such as the CfA (Huchra et al. 1983) observed galaxies
as faint as $m_{Zw}\leq14.5$, while the CfA2 (Falco et al. 1999) is observing
galaxies up to $m_{Zw}\leq15.5$. Approximately 90\% of the sample have been
covered by redshift surveys. To ensure we did not miss any galaxy
without radial velocity information, for those galaxies without known
companions, according to our criteria, we checked NED for possible secondary
galaxies without radial velocity information around the primaries, and also
visually inspected the DSS images for secondary galaxies up to 3 mag fainter
than the primary galaxies. We did not find any possible secondary galaxy
without radial velocity information.

We also checked if the search of galaxies on the DSS was able to detect
low surface brightness galaxies. According to Krongold, Dultzin-Hacyan \&
Marziani (2001) the DSS images have a surface brightness limit of
$\approx22$~mag~pixel$^{-1}$, which, for pixels with a size of
1.7$^{\prime\prime}$, corresponds to $\approx23$~mag~arcsec$^{-2}$.
Most of the low surface brightness galaxies detected by Impey et al. (1996)
have B band central surface brightnesses of the order of $\mu_0>21$ mag
arcsec$^{-2}$ or fainter, effective radii R$_e\sim10^{\prime\prime}$ and
effective surface brightness $\mu_e\sim24$ mag arcsec$^{-2}$. Using the
equations from Impey, Bothum \& Malin (1988), we have that the integrated
magnitude of an exponential disk galaxy can be calculated by
m$_{tot}=\mu_0 - 0.87 - 5 \log(R_e)$, which for a galaxy with $\mu_0=21$
mag arcsec$^{-2}$ and R$_e\sim10^{\prime\prime}$ corresponds to
m$_{tot}=15.13$. Consequently, most of the low surface brightness galaxies
will be fainter than B$=$15.5, which is the faintest magnitude for our
companion galaxies. This means that we did not miss a significant
number of low surface brightness galaxies in the search for companions on the
DSS images.

Table 1 presents the galaxies for which we could find companions. This
Table is separated in 5 parts, according to activity type. It presents the
names of the primaries, the magnitude differences between the primaries and
the secondaries, the distances between them in units of the primary diameter,
the moduli of the difference between the radial velocities and the names of the
secondaries. We also present in this Table the H$\alpha$ luminosities of the
primaries, whether they are ellipticals, S0's or a later type galaxies and
comment if they are members of the Virgo cluster. We consider as late types,
galaxies which have morphological types Sa or later. This morphological
separation is based on the fact that elliptical galaxies have very little
or no internal gas, S0's have some gas, while galaxies of morphological type Sa
and later have gas in their disks which can feed the nucleus or form stars.

We point out the fact that two of the galaxies in Table 1, NGC\,4041 and
NGC\,4564, have a secondary at a distance larger than 5D$_{25}$. We consider
these two galaxies interacting because their secondaries are much larger then
them and their distances are smaller than 5 diameters of the secondary. In
fact, the secondaries also are in the Palomar sample.

One possible problem which could exclude secondary galaxies from our
sample is the case of edge-on galaxies, which can be highly reddened
and thus result in a magnitude difference larger than 3 mag. Low mass
starbursting galaxies can also create a selection bias, since the starburst
would make them brighter and they would be considered as a secondary.
However, these effects are expected to affect galaxies of all activity types,
and given the size of the sample, they should be smeared out, since
adding or subtracting one galaxy with a companion will not change the
statistical results significantly.

Another effect which can influence the results in the case of broad lined
Seyfert galaxies, is the contribution from the nucleus to the integrated
magnitude of their hosts. To correct for this effect we used the observed
nuclear B magnitudes given by Ho \& Peng (2001). We found that this effect
is negligible for most of the galaxies in the sample, with the exception of
NGC4151, where the nucleus increases the integrated magnitude by
$\approx0.3$ mag. We consider that this effect is also minimal for
broad lined LINERs, since their nuclear continuum emission at optical
wavelengths is weak (Ho 1999a).

Figure 1 presents the distribution of the modulus of the difference
between the radial velocities of the primary and secondary galaxies.
We can see that although we considered as companions two galaxies with
$| \Delta \rm Vel | \leq 1000$ km s$^{-1}$, $\approx$70\% of the pairs have
velocity differences smaller than 300 km s$^{-1}$. This number increases
to 90\% for velocity differences smaller than 500 km s$^{-1}$.
Figure 1 also presents the distribution of radial velocity moduli of
the sample of galaxy pairs from Karachentsev (1987). The visual inspection
of the two groups of galaxies shows that they have similar distributions,
indicating that most of the galaxies with
$| \Delta \rm Vel | \leq 1000$ km s$^{-1}$ in the Palomar are real pairs.
The KS test confirms this result, giving a 14\% probability that two samples
drawn from the same parent population would differ as much as these two samples.
 
The total number of galaxies in each one of the subsamples as well
as the number of galaxies with companions, percentage of companions
and corresponding 1$\sigma$ uncertainties are given in Table 2.
In this Table we also show these numbers for the cases when we exclude
the galaxies which are members of the Virgo cluster from the sample,
and the results obtained from spliting the activity groups into subgroups of
galaxies with different morphologies, or H$\alpha$ luminosities.
       
\section{Local galaxy densities}

This Section presents the comparison of the local galaxy densities,
$\rho_{gal}$, of galaxies with different activity types. This measurement,
defined by Tully (1988), gives the density of galaxies brighter
than M$_B=-16$ around the galaxy of interest, thus it is an
estimate of their environments. These values are tabulated by Ho et al. (1997a).

Comparing the different activity classes, without excluding any galaxies,
the KS test shows that, with the exception of absorption line galaxies,
there is no statistically significant difference between their $\rho_{gal}$
distributions. In the case of absorption line galaxies, the KS test shows that
their $\rho_{gal}$ distribution usually is different from that of other types
of galaxies at the 1\% significance level. These results can be seen on the
top left panel of Figure 2, where we show the percentage of galaxies, of
a given activity type, as a function of Log($\rho$).
The analysis of the different distributions shows
that absorption line galaxies are skewed towards higher density regions,
due to the fact that $\approx$50\% of them are in the Virgo
cluster. In fact, once we compare only those galaxies which are not members of
the Virgo cluster (top right panel of Figure 2), there is no statistically
significant difference in the $\rho_{gal}$ distributions. The same result is
obtained if we compare only the galaxies which are members of the Virgo
cluster. 

This test was repeated separating the galaxies according to their
H$\alpha$ luminosities. We found that comparing only the galaxies
with H$\alpha$ luminosities smaller, or larger than 10$^{39}$ erg s$^{-1}$
gives essentially the same result, that the $\rho_{gal}$ distributions
of most types of galaxies are not significantly different. The corresponding
Log($\rho$) percentage distributions are presented in the middle panels of
Figure 2 left for L(H$\alpha$)$<10^{39}$ erg s$^{-1}$, and right for
L(H$\alpha$)$>10^{39}$ erg s$^{-1}$.
Notice that absorption line galaxies are not included in this
comparison because they do not have emission lines. 
Only in the comparison  between LINERs and HII galaxies with
L$_{H\alpha}>10^{39}$ erg s$^{-1}$, we find a difference in the $\rho_{gal}$
distribution, with the KS-test showing a 2\%
probability that two samples drawn from the same parent population would differ
as much as these two samples. This result is due to the fact that these HII
galaxies are found, on average, in slightly lower galaxy density regions
than the LINERs. This result is due to a difference in the morphological
types of these two groups of galaxies, since HII galaxies are found only
in spirals, while a considerable portion of LINERs are in ellipticals,
which are found in higher density environments. We discuss this effect
in more detail in the next Section.

A similar result was obtained when we separated the galaxies by
morphological type, again without a significant difference in the
$\rho_{gal}$ distributions, which can be seen in the bottom left panel
of Figure 2 for Early type galaxies (ellipticals and S0's) and bottom
right panel for Late type galaxies (later than S0a). Only in the comparison
between LINERs and absorption line galaxies in Early type systems,
we find a 1.4\% probability that two samples drawn from the same
population would differ as much as these galaxies.
This result is due to the fact that
almost all absorption line galaxies are Early type systems, approximately
half of them are in the Virgo cluster and consequently have higher local
galaxy densities.

Another test done using the local galaxy density estimator was to
compare, for galaxies of the same activity type, if there was any
difference in the $\rho_{gal}$ distributions when we compare galaxies of
Early and Late morphological types, or for L$_{H\alpha}>10^{39}$ and
$<10^{39}$ erg s$^{-1}$. Again we did not find any significant difference,
indicating that galaxies with the same activity type, but different
morphological types, or luminosities are found in similar environments.

\section{Percentage of companion galaxies}

This section shows the results of the comparison between
the percentage of galaxies with companions among different
activity classes. Besides separating the galaxies according to
their activity types, we also split them based on
membership to the Virgo cluster, morphological type and H$\alpha$ luminosity.

The comparison between the different types of galaxies is done using
contingency table analysis (Press et al. 1993), which is a measure
of association for two distributions. Given the number of galaxies
with and without companions in two of our subsamples (e.g. Seyferts
and LINERs), this analysis  gives the probability of any of these
measurements being correlated.

\subsection{All galaxies}

Figure 3 shows the percentage of galaxies with companions
as a function of the distance between the primary and the secondary, in units
of the primary diameter (D$_{25}$). Although it would be ideal to plot the
uncertainties of the percentage of galaxies with companions in the Figures,
this would result in extremely dense figures, with a large amount of lines
superposing each other, which would be
very difficult to understad. Instead of doing this, we give the percentages,
together with the corresponding 1$\sigma$ uncertainties in Table 2. Each
panel in Figure 3 shows a different range of magnitude differences between
the primary and secondary galaxies. We can see that, besides the fact that the
percentage of galaxies with companions is smaller for smaller magnitude
differences between the primary and the secondary, the distributions do not
differ too much for different $| \Delta m |$ values. The most noticeable
effect in this plot is the fact that for magnitude differences of
$| \Delta m | \leq 1$ mag, there are no Seyfert galaxies with companions.

The analysis of Figure 3 shows that, contrary to what previous papers have
shown, Seyfert and HII galaxies are the ones with the smaller percentage of 
companion galaxies. On the other hand, LINERs, transition and
absorption line galaxies have the larger percentage of companions.
In particular, the frequency of LINERs with companions is almost
twice that of Seyferts and HII galaxies.

The results of the comparisons presented below are summarized in Table 3.
This Table presents the probability that the null hypothesis, that there is
no correlation between the two quantities being compared, is correct.
This probability is calculated based in $\chi^2$ statistics.
We consider that there is an excess of companions in galaxies with a
given activity type relative to another, only if the contingency table
analysis gives a probability smaller than 5\%. This means that the null
hypothesis, that there is no correlation, is rejected at the 2$\sigma$ level.

The contingency table analysis confirms that Seyferts have a smaller
percentage of companions when compared to LINERs, with only  a
3.3\% probability, that the null hypothesis can be accepted.
A similar result is obtained for HII galaxies, which have only a
0.2\% probability of having the same percentage of companions as LINERs,
and 1.6\% when compared to absorption line galaxies. All other combinations
of different subclasses have similar percentages of companion
galaxies within the statistical uncertainties.

We also tested whether considering galaxies as companions only if their
distances are smaller than 3 times the diameter of the primary, the
original Rafanelli et al. (1995) criterion, would give a different result.
We find that, in most of the cases,
it only reduces slightly the probability of two samples
having the same percentage of companion galaxies, from 3.3\% to 0.6\% in
the case of the comparison between Seyferts and LINERs, from 0.2\% to
0.001\% for LINERs and HII galaxies, from 6.5\% to 2.4\% for transition
and HII galaxies, from 7.7\% to 10.5\% for Seyferts and absorption
galaxies, and from 1.6\% to 1\% for HII and absorption
galaxies. This same test was done for all the comparisons presented below, with
similar results for 3 and 5 diameters, so, hereafter we will discuss
only the results obtained when considering distances smaller than 5
diameters of the primary.

\subsection{Excluding galaxies in the Virgo cluster}

As we did in Section 4, for the comparison between the local galaxy density
distributions, here we repeat the analysis presented above excluding
the galaxies which are in the Virgo cluster. We do this because
these galaxies are in a higher density environment, and consequently
have a higher probability of having close companions. Since $\approx$50\%
of the absorption line galaxies are in Virgo, a much higher proportion than
for the other activity classes, this can influence the results.

Figure 4 presents, in a similar way to Figure 3, the percentage of galaxies,
which are not members of the Virgo cluster, with companions. The results
are very similar to those obtained when we do not exclude the Virgo cluster
members (Figure 3). The only difference is that now LINERs do not have a
significant excess of companions when compared to Seyferts (6.5\%),
which is also the case for the comparison between HII and absorption
line galaxies (7.2\%). We still find a significant excess of companions
when we compare LINERs to HII galaxies, with a probability of only 0.04\%
that these two subsamples are similar.

\subsection{Different morphological types}

Another test done to the sample was to separate the different activity types
into subgroups, based on their host galaxy morphological types. We decided
to separate the galaxies into ellipticals, S0's, and late type galaxies,
those with morphological types Sa and later.
This division was based on the fact that late type galaxies have enough
gas in their disks, and an interaction can easily move this gas into
the nuclear region. In the case of elliptical galaxies, the amount of
gas is much smaller, so the source of gas to feed their nuclei may
have to be external to the galaxy. In S0 galaxies the amount
of gas is larger than that of ellipticals, so we test whether combining
these galaxies with ellipticals or late types makes a difference to the
results.

In the comparison between the percentage of companions in elliptical
galaxies of different activity classes, we do not include HII galaxies
and Seyferts, because there are no HII galaxies in Ellipticals and only
a small number of Seyferts with this morphology.
Similarly, when analyzing the late type galaxies we do not include
absorption line objects.

The results of the separation of the sample into ellipticals, late types,
early types (ellipticals plus S0's), and late types plus S0's
is presented in Figure 5. In the case of early types, there
is no significant excess of companions among most of the galaxies studied.
The comparison between the late type
galaxies shows that all different activity classes have similar percentages
of galaxies with companions. When comparing only elliptical galaxies,
we find that there is no significant difference between the different
activity classes. A similar result is obtained when we add late types
and S0's, with the exception that in this case LINERs have a slightly
higher percentage of companions when compared to HII galaxies, but the
probability that the two distributions are similar is 4.8\%, only marginally
significant.

An interesting result which comes out of the analysis of Figure 5, is the
fact that LINERs and transition galaxies in ellipticals have $\sim$2 times
higher a percentage of companions than the corresponding galaxies in late
type systems. In the case of LINERs, the probability of the two samples
being the same is 0.4\%. For transition objects this result is
not statistically significant, given the small number of transition
galaxies in ellipticals. This result is maintained when we compare
early (E$+$S0) and late type LINERs, but in this case the probability
of the two samples being the same is 4\%.

\subsection{Different H$\alpha$ luminosities}

The results of separating the sample into galaxies with H$\alpha$
luminosities smaller or larger than 10$^{39}$ erg s$^{-1}$ is presented
in Figure 6. Absorption line galaxies are not used in this analysis.
For galaxies with L$_{H\alpha}\leq10^{39}$ erg s$^{-1}$, there is no
difference in the percentage of LINERs, Seyferts and transition galaxies
with companions. However, LINERs and transition galaxies have a significantly
higher percentage of companions when compared to HII galaxies, with the
null hypothesis that the there is no correlation between the activity type
and the percentage of companions being rejected at the 0.3\%
and 1.5\% significance levels, respectively. The comparison of Seyferts
and HII galaxies does not show a significant difference (11\%).

This result is again due to the mixture of ellipticals and spirals in the
LINERs and transition galaxies samples. If we consider only spiral galaxies
(S0's + late types)
with L(H$\alpha$)$\leq10^{39}$ erg s$^{-1}$, we find that there is significant
difference between the percentage of companion in these galaxies and
HII galaxies.

The comparison of galaxies with L$_{H\alpha}>10^{39}$ erg s$^{-1}$
shows that there is no statistically significant difference in the
percentage of galaxies with companions among different activity classes.
The comparison between LINERs and Seyferts shows a small probability
that they have similar percentages of companions, 5.9\%, but this can
be due to the small number of Seyferts.

We compared if there is a difference in the proportion of galaxies with
companions for objects with the same activity class, but H$\alpha$
luminosities larger and smaller than 10$^{39}$ erg s$^{-1}$. There is
no significant difference for LINERs, transition galaxies and Seyferts.
However, there is a significantly higher percentage of galaxies with
companions in the case of
HII galaxies with L(H$\alpha$)$>10^{39}$ erg s$^{-1}$, than in HII galaxies
with L(H$\alpha$)$<10^{39}$ erg s$^{-1}$, rejecting the null hypothesis
at the 2.9\% significance level. This result agrees with the ones
presented by Bushouse (1986), Keel et al. (1985) and Kennicutt et al. (1987).
It also agrees with the result obtained by Ho, Filippenko \& Sargent (1997c),
who found, in this sample, that HII galaxies in barred early type hosts
have higher H$\alpha$ luminosities.
In fact, the percentage of HII galaxies with companions increases to
$\approx$65\% for galaxies with L(H$\alpha$)$>10^{40}$ erg s$^{-1}$, a total
of 9 out of 14 galaxies with companions. This percentage is significantly
higher than that of the rest of the HII galaxies, as well as all the other
galaxies in the sample, rejecting the null hypothesis
at the 0.007\% and 1.5\% significance levels, respectively.
 
\section{Discussion and Summary}

The previous section shows that a simple analysis, not taking into
account morphology or activity level (H$\alpha$ luminosity) of the galaxies,
gives that  the percentage of LINERs, transition and absorption
line galaxies with nearby companions is significantly higher than that of
Seyferts and HII galaxies. This result contradicts all previous works in this
field, which either found that Seyferts and HII galaxies have the same
percentage of companions as other galaxies, or that they have a higher
percentage of companions.
However, we show that when we take into account the morphological types of
the galaxies and split them into subgroups containing only ellipticals and only
spirals, the situation is different. There is no difference
in the percentage of galaxies with companions among different
activity types if we consider only galaxies of similar morphological types.
This result is consistent with the one found by Fuentes-Williams \& Stocke
(1988), Bushouse (1986, 1987), De Robertis et al. (1998), and also with
more recent results on clustering of low luminosity AGNs at higher redshifts
(Brown et al. 2001; Schreier et al. 2001). The percentage of elliptical
galaxies with companions is approximately two times higher than that of
spirals. This explains why LINERs, transition and absorption line galaxies
have higher percentages of companions when all galaxies are considered,
since a higher percentage of these galaxies is found in ellipticals when
compared to Seyferts and HII galaxies.

Given the fact that most of the HII galaxies in the Palomar sample have only
small quantities of recent star formation, most of them
should be considered as normal, quiescent galaxies. We found
that a there is a higher percentage of galaxies with companions among
HII galaxies with L(H$\alpha$)$>10^{39}$ erg s$^{-1}$ then in lower
H$\alpha$ luminosity HII galaxies. The percentage of companions increases
even more for HII galaxies with higher H$\alpha$ luminosities,
which confirms previous results (Kennicutt et al. 1997, Keel et al. 1985,
Bushouse 1986).

The results we obtained separating the galaxies by morphological
types is somewhat expected. It was shown by Dressler (1980) and
Whitmore, Gilmore \& Jones (1993) in the study of clusters of galaxies
that the percentage of ellipticals increases in higher density environments.
Although this result was based on clusters, Postman \& Geller (1984) found
that the results also apply for groups of galaxies. Furthermore,
Whitmore et al. (1993) found that the percentage of ellipticals in close
pairs is higher than that of spirals, supporting the results we found.
Charlton, Whitmore \& Gilmore (1995) showed that ellipticals are found more
often in pairs than spirals in clusters and high density environments, like
large groups. We found that $\sim$30\% of the ellipticals in our sample are
in the Virgo cluster, while only 20\% of the spirals are found in Virgo.
Another $\sim$30\% of the ellipticals are found in groups of galaxies
with 10 or more galaxies Garcia (1993), suggesting that the higher percentage
of ellipticals with companions is in fact due to the morphology-density
relation.

The study of the local galaxy densities (Section 4) shows that, in most of
the cases, the distribution of this quantity does not differ for different
activity types. In the cases where we find significant differences, they
can be explained if we exclude the galaxies which belong to the Virgo
cluster, or if we separate the galaxies into different morphological types.
These results agree with the ones obtained in the comparison between the
percentage of companions in galaxies with different activity types.

It can be argued that the results found for Seyfert galaxies is somewhat
questionable, since the Palomar survey has a large portion of low
luminosity Seyfert galaxies, which are usually not observed in other
surveys. In this way, the results presented here would be biased
towards low luminosity Seyferts, which, in a way similar to HII galaxies,
may have a higher percentage of companions as the luminosity increases.
Evidence of this effect is given by Heckman, Carty \& Bothun (1985),
Heckman et al. (1984) and Hutchings (1983), who showed that radio galaxies and
quasars are found in higher density environments. We believe that this is not
the case. We compare the results presented here with the ones from Schmitt
et al. (2001) for warm infrared Seyfert galaxies. The median [OIII] luminosity
of their Seyfert galaxies, which is believed to be an isotropic indicator
of the intrinsic luminosity of these galaxies,
is L([OIII])$=7\times10^{40}$ erg s$^{-1}$, calculated using de Grijp et al.
(1992) values. This value corresponds to 70 times the median value of the
[OIII] luminosity of Seyfert galaxies in the Palomar survey
(L([OIII])$=10^{39}$ erg s$^{-1}$), obtained from Ho et al. (1997a).
Schmitt et al. (2001) observed that between 19\% and 28\% of their
Seyfert galaxies have companions.
Their criteria were a little different from ours, assuming that galaxies were
companions if the distance between the primary and the secondary was smaller
than 3 times the diameter of the primary, and the brightness
difference between them smaller than 3 mag. Since it was not possible
for them to find radial velocities for all possible companion galaxies,
they could only put limits on the percentage of galaxies with companions.
The lower limit is 19\%, which corresponds to those galaxies where it was
possible to confirm that the companion has a radial velocity difference
smaller than 1000 km~s$^{-1}$, while the upper limit is 28\%, corresponding
to all galaxies, including those without information about the velocity
of the secondary.
The uncertainty in these measurements is $\sim$5\%, given by Poisson statistics.
The number of Seyfert galaxies with companions at distances smaller than
3 diameters in the Palomar survey is 8 in a sample of 46 galaxies,
a total of 17\%$\pm6$\%. Given the uncertainties involved in these
measurements, there is no significant difference in the percentage of
Seyfert galaxies with companions in the Palomar and Schmitt et al. (2001)
samples, which shows that low and high luminosity Seyfert galaxies similar
environments. The contingency table analysis gives that there is a 16\%
probability that the two samples have the same number of companions.

The explanation of why other papers found that there is a higher percentage
of Seyferts with companions than normal galaxies with companions is not
very clear. One possibility would be that they mixed galaxies of
different morphological types. However, the percentage of Seyferts in
ellipticals is small and this effect could be smeared out in a larger
sample. Another explanation is related to the way the previous works selected
their samples of Seyferts. In the case of Laurikainen \& Salo (1995)
and Dultzin-Hacyan et al. 1994), their samples included a large portion of
Seyferts selected from ultraviolet surveys. In Seyfert 2 galaxies, the
Unified Model predicts that the ultraviolet emission is either nuclear
radiation reflected towards the observer (Antonucci 1993), or a circumnuclear
starburst (Cid Fernandes \& Terlevich 1995). If the Seyfert 1's and Seyfert 2's
have the same amount of ultraviolet excess, this
means that the Seyfert 2's were selected from two orders of magnitude
higher in the luminosity function, since only 1\% of the nuclear light is
believed to be reflected, or that their circumnuclear region is dominated
by a luminous starburst which is responsible for most of the ultraviolet
emission (see Schmitt et al. 2001 for a discussion about this).
In the latter case, the fact that these papers
observed a higher percentage of Seyfert 2's with companions, but not Seyfert
1's, would be due to the fact that the percentage of starbursts with
companions goes up as the starburst luminosity goes up, as shown above.

The result that Seyfert galaxies have the same percentage of companion
galaxies as galaxies with other activity types is intriguing and may
have important consequences for the theories of how the gas is moved from
kiloparsec scales to the inner parsec region of the galaxy and feeds the nuclear
black hole. Taken at face value, this result indicates that interactions are
not important in this process. However, this result can be interpreted in a
different way. Since we expect a delay between the time when
the interaction occurs and when the gas reaches the nucleus,
we may be seeing different stages of this process. It has been shown by
several simulations (e.g. Byrd et al. 1986, 1987, Byrd \& Valtonen 2001, Lin
et al. 1988, Hernquist \& Mihos 1995) that interactions move the gas to
the nuclear region of the galaxy, where its density and temperature increases,
a starburst is formed and later the gas feeds the black hole.
Also, taking into account the evidence that many galaxies
have black holes in their nuclei, suggesting that this may be the case
in all galaxies (Magorrian et al. 1998, Ho 1999b, Gebhardt et al. 2000,
Ferrarese \& Merritt 2000), it may be possible that all galaxies
pass through a period of activity. The galaxies may also pass through
different activity types, where they first present a star formation period,
when the gas has just moved into the nuclear region, later they present a
period of Seyfert activity, when the gas is being accreted by the black hole,
and finally they go into a period where they turn into a LINER or transition
galaxy, when the amount of gas available to fuel the nucleus is small.

\acknowledgements

We would like to thank Jim Ulvestad, Luis Ho and the referee
for comments and suggestions
which significantly improved this manuscript. This research made use of the
NASA/IPAC Extragalactic Database (NED), which is operated by the Jet Propulsion
Laboratory, Caltech, under contract with NASA. We also used the Digitized Sky
Survey, which was produced at the Space telescope Science Institute under U.S.
Government grant NAGW-2166.  The National Radio Astronomy Observatory is a
facility of the National Science Foundation operated under cooperative
agreement by Associated Universities, Inc.

\begin{figure}
\psfig{figure=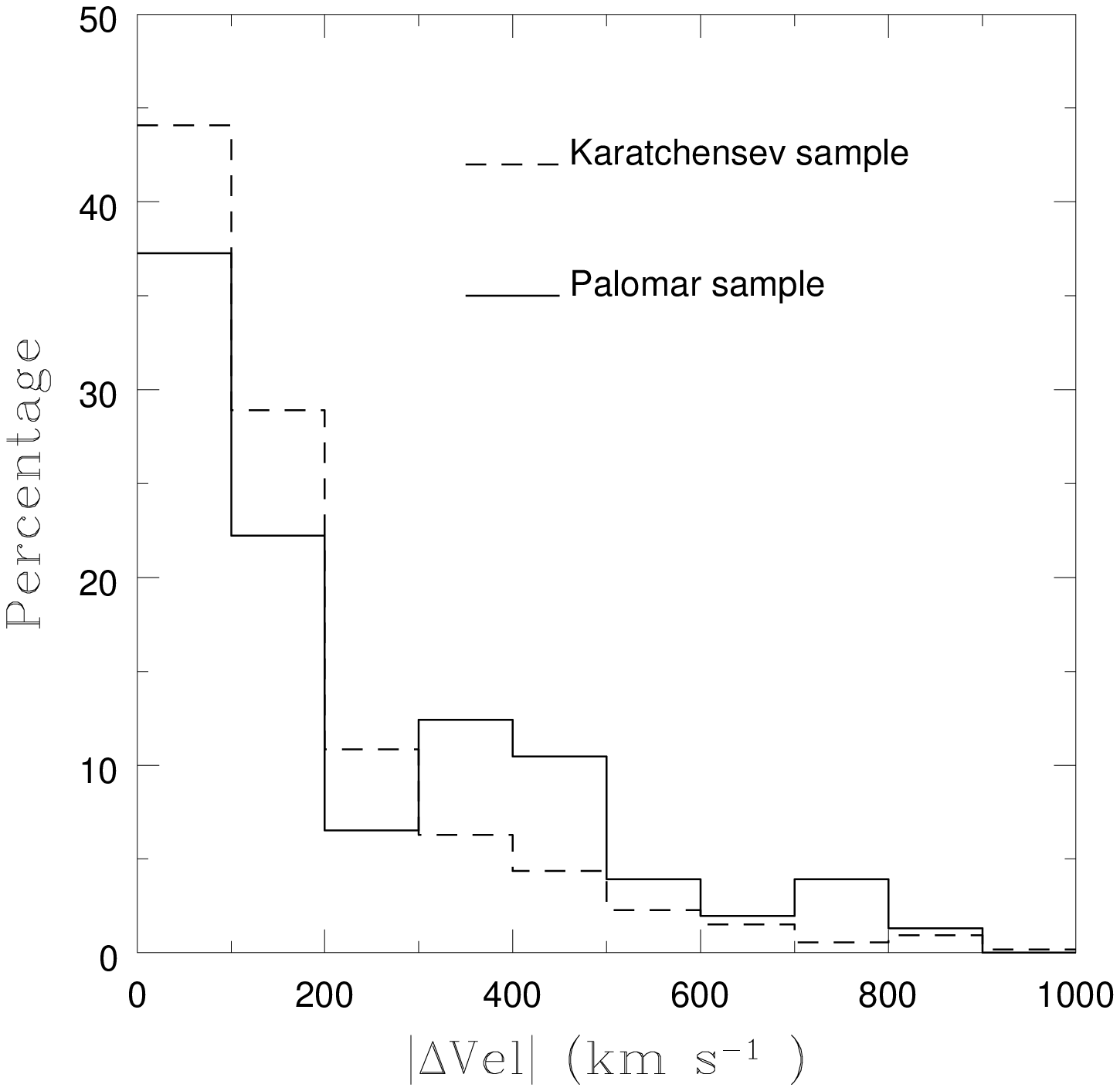,width=10cm,height=10cm}
\caption{Percentage distribution of galaxies as a function of the
modulus of the difference between the radial velocity of the primary and
secondary galaxies. The solid line represents the Palomar sample, while
the dotted line represents the Karachentsev (1987) sample of pair galaxies.}
\end{figure}

\begin{figure}
\vskip 1cm
\psfig{figure=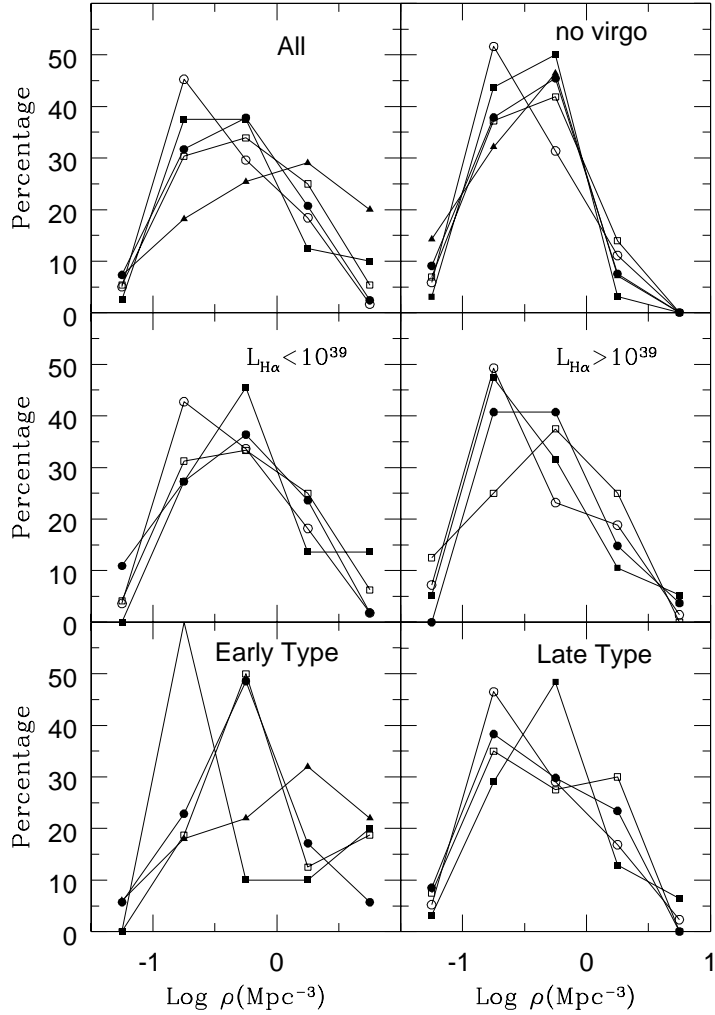,width=15cm,height=15cm}
\caption{Percentage of galaxies of different activity types as a function of
the Logarithm of the local galaxy density ($\rho$), binned in intervalas of
Log($\rho$)=0.5. LINERs are represented by filled circles,
absorption line galaxies by filled triangles, transition galaxies
by open squares, Seyferts by filled squares and HII galaxies
by open circles. The histogram for all galaxies is presented on the top left
panel; all galaxies excluding the ones in the Virgo cluster on the top right
panel; galaxies with L(H$\alpha$)$<10^{39}$ erg s$^{-1}$ middle left;
 galaxies with L(H$\alpha$)$>10^{39}$ erg s$^{-1}$ middle right; Early type
galaxies (Ellipticals and S0's) bottom left; Late type galaxies (later than
S0a) bottom right.}
\end{figure}

\begin{figure}
\vskip -5cm
\psfig{figure=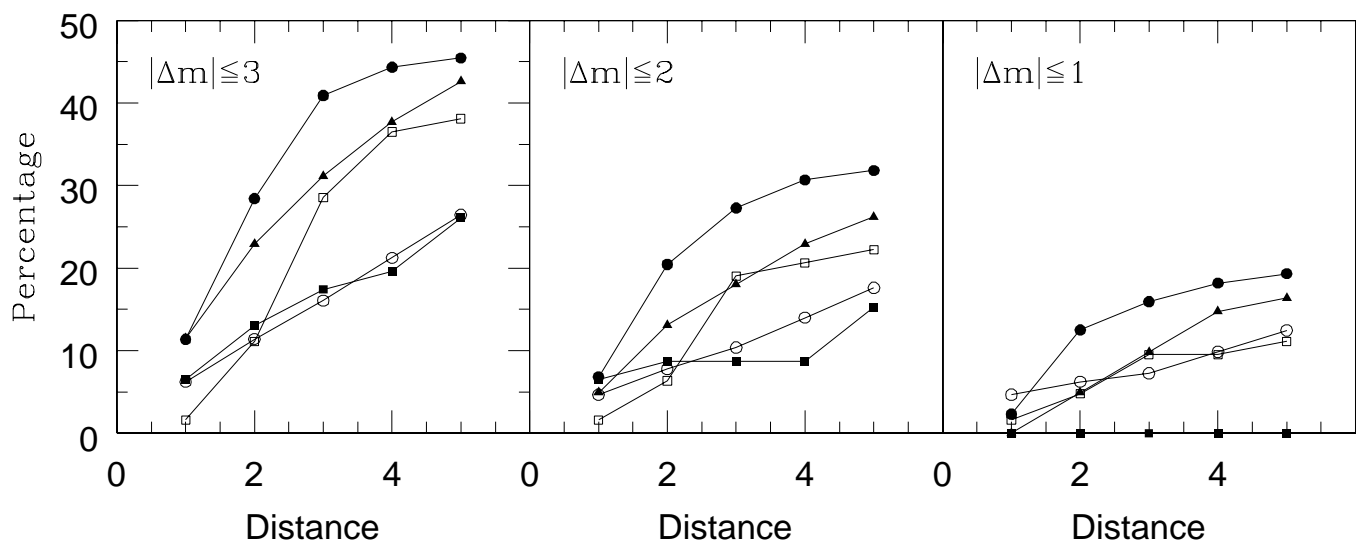,width=20cm,height=20cm}
\caption{The percentage of galaxies with companions as a function
of the distance between the two galaxies, where the distance is
measured in units of the diameter (D$_{25}$) of the primary galaxy.
The left panel shows the results for galaxies with magnitude differences
smaller than $| \Delta m | \leq 3$ mag, the middle panel for
galaxies with magnitude differences smaller than $| \Delta m | \leq 2$ mag,
and the right panel for magnitude differences smaller than
$| \Delta m | \leq 1$ mag. Symbols as in Figure 2.}
\end{figure}
  
\begin{figure}
\psfig{figure=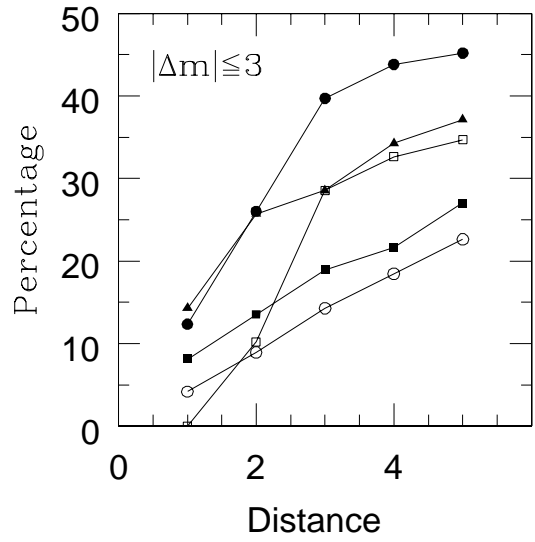,width=20cm,height=20cm}
\caption{Same as Figure 2, but excluding the galaxies in the Virgo cluster.
We show only the $| \Delta m | \leq 3$ mag because the other magnitude cuts
have similar distributions.}
\end{figure}
  
\begin{figure}
\psfig{figure=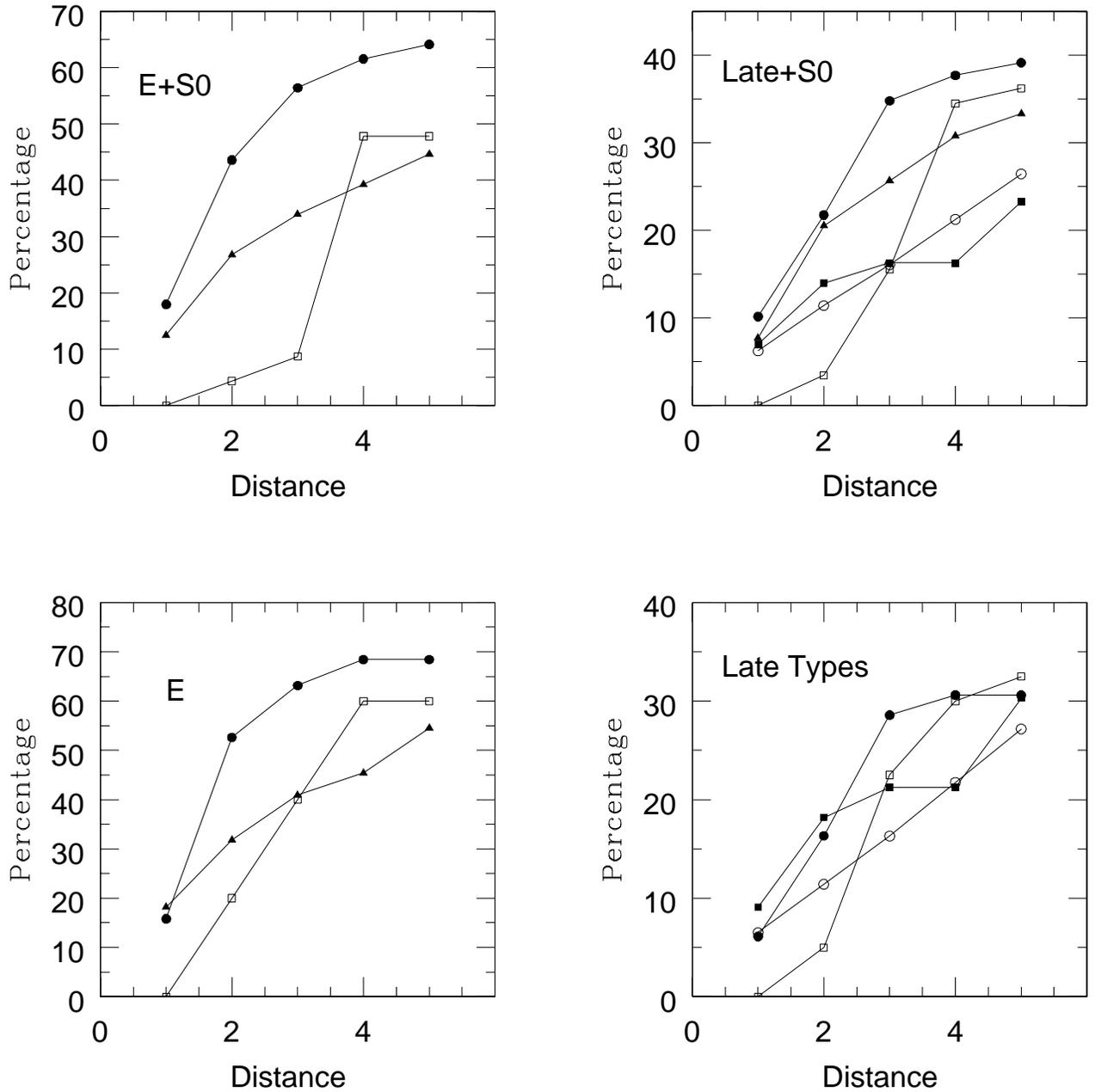,width=20cm,height=20cm}
\caption{Same as Figure 2, but separating between galaxies with different
morphological types. Ellipticals (bottom left), ellipticals plus S0's (top
left), galaxies with morphological type Sa and later (bottom right), and
the sum of late type galaxies and S0's (top right). We do not exclude the
members of the Virgo cluster.}
\end{figure}
  
\begin{figure}
\psfig{figure=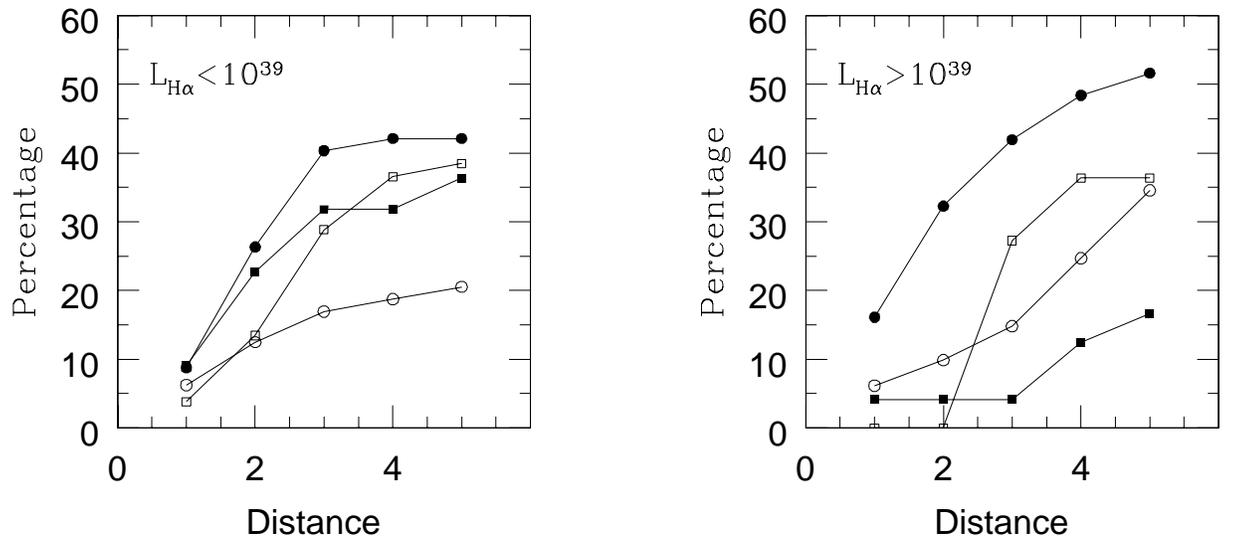,width=20cm,height=20cm}
\caption{Same as Figure 2, but separating between galaxies with H$\alpha$
luminosities smaller than 10$^{39}$ erg s$^{-1}$ (left), and those with
H$\alpha$ luminosities higher than 10$^{39}$ erg s$^{-1}$ (right).
We do not separate the galaxies by morphological type, or exclude the
ones in the Virgo Cluster.}
\end{figure}

\begin{tiny}
\begin{deluxetable}{lrrrlrrrl}
\tablewidth{0pc}
\tablecaption{Sample Properties}
\tablehead{\colhead{Primary}&\colhead{$\Delta$B}&\colhead{Distance}&
\colhead{$|\Delta Vel|$}&\colhead{Secondary}&\colhead{L$_{H\alpha}$}&
\colhead{T}&\colhead{Virgo}&\colhead{Comments}\cr
&(mag)&(Diam.)&(km s$^{-1}$)&&(erg s$^{-1}$)&&&\cr
(1)&(2)&(3)&(4)&(5)&(6)&(7)&(8)&(9)}
\startdata
\multicolumn{8}{c}{}\cr
\multicolumn{8}{c}{Absorption Line Galaxies}\cr
\multicolumn{8}{c}{}\cr
\tableline
NGC\,0507 & -1.70 & 0.46 & 602 & NGC\,0508  &  ---  & S0&     &   \cr
NGC\,1023 & -2.87 & 0.33 & 111 & NGC\,1023\,A& ---  & S0&     &   \cr
NGC\,2300 &  0.02 & 2.19 & 454 & NGC\,2276  &  ---  & S0&     &   \cr
NGC\,2950 & -2.83 & 3.15 &  21 & UGC\,5179  &  ---  & S0&     &   \cr
NGC\,3384 &  0.57 & 1.46 & 154 & NGC\,3379  &  ---  & S0&     &   \cr
NGC\,3613 & -0.69 & 4.97 & 512 & NGC\,3619  &  ---  & E &     &   \cr
NGC\,3640 & -2.78 & 0.64 & 441 & NGC\,3641  &  ---  & E &     &   \cr
NGC\,4262 & -2.09 & 4.92 &  29 & IC\,0781   &  ---  & S0& Yes &   \cr
NGC\,4291 & -0.30 & 3.41 & 399 & NGC\,4319  &  ---  & E &     &   \cr
NGC\,4340 &  0.08 & 1.68 & 326 & NGC\,4350  &  ---  & S0& Yes &   \cr
NGC\,4365 & -2.91 & 1.60 & 444 & NGC\,4370  &  ---  & E & Yes &   \cr
NGC\,4382 & -1.52 & 1.12 & 200 & NGC\,4394  &  ---  & S0& Yes &   \cr
NGC\,4406 & -1.96 & 1.21 & 480 & NGC\,4402  &  ---  & E & Yes &   \cr
NGC\,4417 &  0.07 & 3.76 & 422 & NGC\,4424  &  ---  & S0& Yes &   \cr
NGC\,4473 & -0.27 & 3.37 & 884 & NGC\,4477  &  ---  & E & Yes &   \cr
NGC\,4478 & -0.67 & 2.47 & 596 & NGC\,4476  &  ---  & E & Yes &   \cr
NGC\,4503 & -2.56 & 2.12 & 136 & IC\,3470   &  ---  & L & Yes &   \cr
NGC\,4564 &  1.67 & 10.5$^a$ & 379 & NGC\,4579  &  ---  & E & Yes &   \cr
NGC\,4638 & -2.76 & 0.81 &  16 & NGC\,4637  &  ---  & S0& Yes &   \cr
NGC\,4649 & -2.11 & 0.35 & 302 & NGC\,4647  &  ---  & E & Yes &   \cr
NGC\,4754 &  0.32 & 2.57 & 412 & NGC\,4762  &  ---  & S0& Yes &   \cr
NGC\,5576 & -1.42 & 0.91 & 171 & NGC\,5574  &  ---  & E &     &   \cr
NGC\,5638 & -1.53 & 0.76 &  68 & NGC\,5636  &  ---  & E &     &   \cr
NGC\,7332 & -0.15 & 1.50 &  44 & NGC\,7339  &  ---  & S0&     &   \cr
NGC\,7457 & -2.30 & 1.87 &  96 & UGC\,12311 &  ---  & S0&     &   \cr
NGC\,7619 & -2.60 & 1.06 & 369 & NGC\,7617  &  ---  & E &     &   \cr
\tableline
\multicolumn{8}{c}{}\cr
\multicolumn{8}{c}{LINERs}\cr
\multicolumn{8}{c}{}\cr
\tableline
NGC\,0315 & -1.72 & 1.81 & 130 & NGC\,0311  & 39.55 & E &     &    \cr
NGC\,0474 & -0.04 & 0.79 &  11 & NGC\,0470  & 38.52 & S0&     &    \cr
NGC\,1961 & -2.76 & 2.71 &  42 & UGC\,3342  & 39.81 & L &     &    \cr
NGC\,2336 & -1.61 & 2.73 & 158 & IC\,0467   & 38.39 & L &     &    \cr
NGC\,3166 & -2.87 & 1.00 &   4 & NGC\,3165  & 39.10 & L &     &    \cr
NGC\,3169 & -0.05 & 1.76 & 110 & NGC\,3166  & 39.02 & L &     &    \cr
NGC\,3190 & -1.67 & 1.07 & 253 & NGC\,3187  & 38.82 & L &     &    \cr
NGC\,3193 &  0.29 & 1.92 & 108 & NGC\,3190  & 38.20 & E &     &    \cr
NGC\,3226 &  1.27 & 0.73 & 165 & NGC\,3227  & 38.93 & E &     &Pair   \cr
NGC\,3379 & -0.57 & 1.39 & 185 & NGC\,3384  & 37.94 & E &     &   \cr
NGC\,3414 & -2.41 & 2.24 & 220 & UGC\,5958  & 39.23 & S0&     &   \cr
NGC\,3507 & -0.65 & 3.78 & 152 & NGC\,3501  & 39.39 & L &     &   \cr
NGC\,3607 & -2.21 & 0.64 & 266 & NGC\,3605  & 38.93 & S0&     &   \cr
NGC\,3608 &  0.90 & 1.79 & 189 & NGC\,3607  & 38.28 & E &     &   \cr
NGC\,3623 &  0.52 & 2.07 &  80 & NGC\,3627  & 37.77 & L &     &   \cr
NGC\,3718 & -0.72 & 1.44 &  31 & NGC\,3729  & 38.46 & L &     &   \cr
NGC\,3998 & -1.97 & 1.13 & 353 & NGC\,3990  & 40.00 & S0&     &   \cr
NGC\,4036 & -0.27 & 4.05 & 164 & NGC\,4041  & 39.35 & S0&     &   \cr
NGC\,4111 & -2.43 & 2.40 & 136 & NGC\,4117  & 39.40 & S0&     &   \cr
NGC\,4261 & -2.47 & 0.87 & 322 & NGC\,4264  & 39.35 & E & Yes &   \cr
NGC\,4278 & -1.88 & 0.84 & 410 & NGC\,4283  & 39.17 & E &     &   \cr
NGC\,4374 & -2.87 & 1.59 & 395 & NGC\,4387  & 38.89 & E & Yes &   \cr
NGC\,4394 &  1.52 & 2.04 & 198 & NGC\,4382  & 38.33 & L & Yes &   \cr
NGC\,4438 & -1.12 & 0.53 & 717 & NGC\,4435  & 39.37 & L & Yes &   \cr
NGC\,4486 & -2.72 & 1.10 &  67 & NGC\,4478  & 39.44 & E & Yes &   \cr
NGC\,4550 & -0.41 & 1.10 & 785 & NGC\,4551  & 38.41 & S0& Yes &   \cr
NGC\,4762 & -0.32 & 1.58 & 416 & NGC\,4754  & 37.49 & S0& Yes &   \cr
NGC\,5077 & -0.91 & 1.61 &  39 & NGC\,5079  & 39.67 & E & Yes &   \cr
NGC\,5195 &  1.71 & 0.78 &  22 & NGC\,5194  & 37.94 & S0&     &Pair   \cr
NGC\,5353 & -0.36 & 0.62 & 352 & NGC\,5354  & 39.12 & S0&     &   \cr
NGC\,5363 &  0.08 & 3.56 & 103 & NGC\,5364  & 39.42 & S0&     &Pair   \cr
NGC\,5485 & -1.26 & 2.79 & 594 & NGC\,5486  & 38.35 & S0&     &   \cr
NGC\,5566 & -2.93 & 0.64 & 267 & NGC\,5569  & 38.66 & E &     &   \cr
NGC\,5813 & -2.81 & 2.99 & 395 & NGC\,5811  & 38.56 & E &     &   \cr
NGC\,5850 &  0.48 & 2.36 & 770 & NGC\,5846  & 38.66 & L &     &   \cr
NGC\,5970 & -2.97 & 2.81 &  52 & IC\,1131   & 38.06 & L &     &   \cr
NGC\,5982 &  0.67 & 3.08 & 387 & NGC\,5985  & 38.46 & E &     &Pair   \cr
NGC\,5985 & -0.67 & 1.35 & 386 & NGC\,5982  & 38.94 & L &     &Pair   \cr
NGC\,6340 & -2.56 & 1.89 &  39 & IC\,1251   & 38.50 & L &     &   \cr
NGC\,7626 &  0.13 & 2.57 & 381 & NGC\,7619  & 38.81 & E &     &   \tablebreak
\multicolumn{8}{c}{}\cr
\multicolumn{8}{c}{HII Galaxies}\cr
\multicolumn{8}{c}{}\cr
\tableline
NGC\,0520 &  0.00 & 0.02 &   9 & PAIR       & 39.33 & L &     &Merging   \cr
NGC\,0672 & -0.83 & 1.07 &  83 & IC\,1727   & 37.81 & L &     &Pair   \cr
NGC\,0697 & -2.03 & 4.61 & 167 & NGC\,0694  & 39.01 & L &     &   \cr
NGC\,0812 & -2.99 & 3.64 & 647 & UGC\,1585  & 40.08 & L &     &   \cr
NGC\,0877 & -0.99 & 4.59 & 173 & NGC\,0871  & 39.18 & L &     &   \cr
NGC\,2146 & -2.01 & 2.94 & 604 & NGC\,2146\,A&39.76 & L &     &   \cr
NGC\,2342 & -0.92 & 1.66 &  44 & NGC\,2341  & 40.93 & L &     &   \cr
NGC\,2276 & -0.02 & 2.14 & 454 & NGC\,2300  & 39.87 & L &     &   \cr
NGC\,2750 & -0.61 & 0.36 &   2 & MCG+04-22-012&40.60& L &     &   \cr
NGC\,2964 & -1.21 & 2.10 & 246 & NGC\,2968  & 40.01 & L &     &   \cr
NGC\,3034 &  1.47 & 3.29 & 250 & NGC\,3031  & 39.71 & L &     &   \cr
NGC\,3338 & -2.42 & 3.30 &  85 & UGC\,5832  & 38.15 & L &     &   \cr
NGC\,3389 &  1.08 & 2.33 & 595 & NGC\,3384  & 38.60 & L &     &   \cr
NGC\,3395 & -0.23 & 0.57 &   7 & NGC\,3396  & 39.40 & L &     &Pair   \cr
NGC\,3430 & -0.66 & 1.56 &  90 & NGC\,3424  & 39.04 & L &     &   \cr
NGC\,3504 & -1.36 & 4.46 & 162 & NGC\,3512  & 40.81 & L &     &   \cr
NGC\,3646 & -2.84 & 2.01 & 731 & NGC\,3649  & 39.42 & L &     &   \cr
NGC\,3684 &  0.19 & 4.53 &   9 & NGC\,3686  & 38.67 & L &     &   \cr
NGC\,3686 & -0.19 & 4.32 &   9 & NGC\,3684  & 39.80 & L &     &   \cr
NGC\,3690 &  0.00 & 0.05 &  31 & PAIR       & 40.62 & L &     &Merging   \cr
NGC\,3729 &  0.72 & 4.15 &  30 & NGC\,3718  & 39.41 & L &     &   \cr
NGC\,3963 & -0.98 & 3.02 & 181 & NGC\,3958  & 39.61 & L &     &   \cr
NGC\,4041 &  0.27 & 5.72$^a$ & 164 & NGC\,4036  & 39.46 & L &     &   \cr
NGC\,4088 & -1.58 & 1.97 &   7 & NGC\,4085  & 39.00 & L &     &   \cr
NGC\,4123 & -0.33 & 3.20 &  19 & NGC\,4116  & 40.35 & L &     &   \cr
NGC\,4274 & -2.27 & 2.80 & 129 & NGC\,4283  & 38.47 & L &     &   \cr
NGC\,4273 & -2.36 & 0.81 & 114 & NGC\,4277  & 40.26 & L &     &   \cr
NGC\,4298 &  0.26 & 0.73 &   9 & NGC\,4302  & 39.00 & L & Yes &   \cr
NGC\,4380 & -2.58 & 2.62 &   2 & IC\,3328   & 38.02 & L & Yes &   \cr
NGC\,4424 & -0.07 & 3.06 & 412 & NGC\,4417  & 39.02 & L & Yes &   \cr
NGC\,4469 & -2.54 & 4.20 &  15 & UGC\,7596  & 38.16 & L & Yes &   \cr
NGC\,4477 & -2.08 & 1.40 & 472 & NGC\,4479  & 38.84 & S0& Yes &   \cr
NGC\,4485 &  2.41 & 1.53 &  83 & NGC\,4490  & 37.23 & L &     &   \cr
NGC\,4490 & -2.41 & 0.55 &  83 & NGC\,4485  & 37.78 & L &     &   \cr
NGC\,4496\,A&0.00 & 0.23 &   6 & NGC\,4496  & 38.47 & L & Yes &Merging   \cr
NGC\,4517 & -2.70 & 1.63 & 410 & NGC\,4517\,A&37.37 & L &     &   \cr
NGC\,4532 & -2.66 & 4.26 &  31 & HOLMBERG VII&39.69 & L & Yes &   \cr
NGC\,4535 & -1.83 & 4.25 & 736 & NGC\,4519  & 39.72 & L & Yes &   \cr
NGC\,4536 & -2.86 & 1.11 &  51 & NGC\,4533  & 39.38 & L & Yes &   \cr
NGC\,4567 &  0.61 & 0.44 &  11 & NGC\,4568  & 38.78 & L & Yes &Pair   \cr
NGC\,4568 & -0.61 & 0.28 &  11 & NGC\,4567  & 38.95 & L & Yes &Pair   \cr
NGC\,4618 & -1.72 & 1.99 &  66 & NGC\,4625  & 38.16 & L &     &   \cr
NGC\,4631 & -1.49 & 2.08 &  38 & NGC\,4656  & 37.42 & L &     &   \cr
NGC\,4647 &  2.11 & 0.87 & 302 & NGC\,4649  & 38.52 & L & Yes &   \cr
NGC\,4654 & -1.10 & 3.57 &  29 & NGC\,4639  & 39.07 & L & Yes &   \cr
NGC\,4656 &  0.00 & 0.24 &  27 & NGC\,4657  & 37.95 & L &     &Merging   \cr
NGC\,5112 & -0.94 & 3.34 &  24 & NGC\,5107  & 38.19 & L &     &   \cr
NGC\,5364 & -0.08 & 2.10 & 103 & NGC\,5363  & 38.14 & L &     &   \cr
NGC\,5775 & -1.25 & 1.01 & 111 & NGC\,5774  & 38.76 & L &     &   \cr
NGC\,5905 & -0.07 & 3.27 &  82 & NGC\,5908  & 40.25 & L &     &   \cr
NGC\,5962 & -2.50 & 3.34 &  42 & UGC\,9925  & 39.21 & L &     &   \cr
\tableline
\multicolumn{8}{c}{}\cr
\multicolumn{8}{c}{Seyfert Galaxies}\cr
\multicolumn{8}{c}{}\cr
\tableline
NGC\,0777 & -2.05 & 2.86 & 448 & NGC\,0778  & 38.73 & E &     &   \cr
NGC\,1068 & -1.25 & 4.34 & 140 & NGC\,1055  & 41.55 & L &     &   \cr
NGC\,3031 & -1.47 & 1.34 & 240 & NGC\,3034  & 37.64 & L &     &   \cr
NGC\,3227 & -1.27 & 0.40 & 165 & NGC\,3226  & 40.38 & L &     &Pair   \cr
NGC\,3735 & -2.60 & 3.96 &  87 & UGC\,6532  & 39.82 & L &     &   \cr
NGC\,4168 & -2.07 & 1.06 & 424 & NGC\,4165  & 37.60 & E & Yes &   \cr
NGC\,4258 & -2.79 & 1.83 & 578 & NGC\,4217  & 38.35 & L &     &   \cr
NGC\,4565 & -1.45 & 4.28 & 126 & NGC\,4494  & 37.97 & L &     &   \cr
NGC\,4579 & -1.67 & 4.89 & 379 & NGC\,4564  & 39.44 & L & Yes &   \cr
NGC\,4725 & -2.37 & 2.26 &  16 & NGC\,4747  & 38.19 & L &     &   \cr
NGC\,5194 & -1.71 & 0.40 &  22 & NGC\,5195  & 38.88 & L &     &Pair   \cr
NGC\,5395 & -1.27 & 0.66 &  15 & NGC\,5394  & 38.87 & L &     &   \tablebreak
\multicolumn{8}{c}{}\cr
\multicolumn{8}{c}{Transition Galaxies}\cr
\multicolumn{8}{c}{}\cr
\tableline
NGC\,0410 & -1.31 & 2.24 & 352 & NGC\,0407  & 39.43 & E &     &   \cr
NGC\,0521 & -1.30 & 2.53 & 411 & IC\,1694   & 39.16 & L &     &   \cr
NGC\,0524 & -2.80 & 3.48 &  11 & NGC\,0516  & 38.57 & S0&     &   \cr
NGC\,0660 & -1.47 & 2.55 &  78 & UGC\,1195  & 38.89 & L &     &   \cr
IC\,1727 &   0.83 & 1.12 &  83 & NGC\,0672  & 37.69 & L &     &Pair \cr
NGC\,1055 &  1.25 & 3.96 & 142 & NGC\,1068  & 37.92 & L &     &   \cr
NGC\,1161 & -0.71 & 1.00 & 588 & NGC\,1160  & 38.70 & S0&     &   \cr
NGC\,3245 & -1.59 & 2.91 &  26 & NGC\,3245\,A&39.59 & S0&     &   \cr
NGC\,3627 & -0.52 & 2.17 &  80 & NGC\,3623  & 38.50 & L &     &   \cr
NGC\,3628 & -0.34 & 2.41 &  39 & NGC\,3623  & 36.87 & L &     &   \cr
NGC\,3917 & -2.77 & 2.22 & 133 & NGC\,3917\,A&37.29 & L &     &   \cr
NGC\,4145 & -2.70 & 2.16 & 154 & UGC\,7175  & 37.72 & L &     &   \cr
NGC\,4216 & -2.75 & 1.45 & 101 & NGC\,4222  & 38.53 & L & Yes &   \cr
NGC\,4220 & -0.99 & 4.68 & 250 & NGC\,4218  & 38.26 & L &     &   \cr
NGC\,4281 & -2.25 & 1.93 & 186 & NGC\,4277  & 38.61 & S0&     &   \cr
NGC\,4350 & -0.08 & 2.07 & 319 & NGC\,4340  & 38.26 & S0& Yes &   \cr
NGC\,4435 &  1.12 & 1.64 & 710 & NGC\,4438  & 38.98 & S0& Yes &   \cr
NGC\,4459 & -2.64 & 2.51 & 293 & NGC\,4468  & 38.75 & S0& Yes &   \cr
NGC\,4527 & -2.78 & 3.15 &  19 & NGC\,4533  & 38.86 & L & Yes &   \cr
NGC\,4552 & -2.20 & 3.33 & 862 & NGC\,4551  & 38.52 & E & Yes &   \cr
NGC\,4569 & -2.53 & 3.87 & 430 & NGC\,4531  & 39.91 & L & Yes &   \cr
NGC\,5354 &  0.36 & 0.85 & 352 & NGC\,5353  & 38.71 & S0&     &    \cr
NGC\,5746 & -1.87 & 2.36 & 147 & NGC\,5740  & 38.48 & L &     &    \cr
NGC\,5846 & -2.58 & 1.75 & 324 & NGC\,5845  & 38.81 & E &     &    \cr
\tablenotetext{}{Column 1: name of the primary galaxy; Column 2: B magnitude
difference between the primary and the secondary; Column 3: distance
between the two galaxies in units of the diameter of the principal
galaxy; Column 4: The modulus of the difference between the velocities
of the galaxies; Column 5: Name of the secondary galaxy; Column 6: H$\alpha$
luminosity of the principal galaxy; Column 7 indicates the galaxy
morphology E for Ellipticals, S0 for S0's and S0/a's, and, L for Sa
and later. Column 8 indicates if the galaxy
is a member of the Virgo Cluster; Column 9 gives comments.}
\tablenotetext{a}{These two galaxies were considered as interecting
eventhough their distances were larger than 5 diameters. We did this
because their companion galaxies had much larger diameters and their
distances were smaller than 5 diameters if we used the diameter of
the secondary galaxy.}
\enddata
\end{deluxetable}                                                              

\begin{deluxetable}{lrrrrrrrrrr}
\tablewidth{0pc}
\tablecaption{Numbers of galaxies and galaxies with companions in each subsample}
\tablehead{Sample separation&
\multicolumn{2}{l}{Seyferts}&
\multicolumn{2}{l}{HII Galaxies}&
\multicolumn{2}{l}{LINERs}&
\multicolumn{2}{l}{Transition}&
\multicolumn{2}{l}{Absorption}\cr
&Total& Companions&Total& Companions&Total& Companions&Total& Companions&
Total& Companions}
\startdata
All galaxies                & 46&12 (26\%$\pm$8\%)&193&51 (26\%$\pm$4\%)&
88&40 (45\%$\pm$7\%)&63&24 (38\%$\pm$8\%)&61&26 (43\%$\pm$8\%)\cr
Excluding Virgo             & 37&10 (27\%$\pm$9\%)&168&38 (23\%$\pm$4\%)&
73&33 (45\%$\pm$8\%)&49&17 (35\%$\pm$8\%)&35&13 (37\%$\pm$10\%)\cr
Ellipticals                 &  3& 2 (67\%$\pm$47\%)&  0& 0&
19&13 (68\%$\pm$19\%)& 5& 3 (60\%$\pm$35\%)&22&12 (55\%$\pm$16\%)\cr
S0's                        & 10& 0&  9& 1 (11\%$\pm$11\%)&
20&12 (60\%$\pm$17\%)&18& 8 (44\%$\pm$16\%)&34&13 (38\%$\pm$11\%)\cr
Late Types                  & 33&10 (29\%$\pm$10\%)&184&50 (27\%$\pm$4\%)&
49&15 (31\%$\pm$8\%)&40&13 (33\%$\pm$9\%)&5&1 (20\%$\pm$20\%)\cr
L$_{H\alpha}<10^{39}$       & 22& 8 (36\%$\pm$13\%)&112&23 (21\%$\pm$4\%)&
57&24 (42\%$\pm$9\%)&52&20 (38\%$\pm$9\%) &0 &0 \cr
L$_{H\alpha}>10^{39}$       & 24& 4 (17\%$\pm$8\%)& 81&28 (35\%$\pm$7\%)&
31&16 (52\%$\pm$13\%)&11&4 (36\%$\pm$18\%)& 0& 0\cr
\tablenotetext{}{Column 1: the way the sample was separated into subsamples;
Column 2  for Seyferts,  4 for HII galaxies, 6 for LINERs, 8 for Transition
galaxies, and 10 for Absorption line galaxies: the total number of galaxies
of a give activity type in each one of the subgroups; Column 3 for Seyferts,
5 for HII galaxies, 7 for LINERs, 9 for Transition galaxies and 11 for
Absorption line galaxies: the number of galaxies of a given activity type
with companions in each one of the subgroups. We give inside parenthesis
the corresponding percentage of companions and the the 1$\sigma$ uncertainty
in this measurement, calculated using Poisson statistics.}
\enddata
\end{deluxetable}

\begin{deluxetable}{lrrrrrrrrrr}
\tablewidth{0pc}
\tablecaption{Results of the contingency table analysis}
\tablehead{\colhead{Sample separation}&\colhead{L$\times$T}&
\colhead{L$\times$S}&\colhead{L$\times$H}&\colhead{L$\times$A}&
\colhead{T$\times$S}&\colhead{T$\times$H}&\colhead{T$\times$A}&
\colhead{S$\times$H}&\colhead{S$\times$A}&\colhead{H$\times$A}}
\startdata
All galaxies&44.5\%&3.3\%&0.2\%&77.8\%&16.9\%&6.5\%&65.9\%&96.3\%&7.7\%&1.6\%\cr
All galaxies distance $<$3 Diameters&13.6\%&0.6\%&$<$0.1\%&22.5\%&16.2\%&2.4\%&79.8\%&82.6\%&10.5\%&1.0\%\cr
Excluding Virgo&24.7\%&6.5\%&0.04\%&42.8\%&44.8\%&8.7\%&81.7\%&56.7\%&35.8\%&7.2\%\cr
Ellipticals&72.2\%&---&---&36.4\%&---&---&82.5\%&---&---&---\cr
Ellipticals $+$ S0s&21.0\%&---&---&6.2\%&---&---&79.6\%&---&---&---\cr
Late Types&73.5\%&97.6\%&66.4\%&14.5\%&84.0\%&52.3\%&13.0\%&73.8\%&15.1\%&17.0\%\cr
Late Types $+$ S0s&73.5\%&8.2\%&4.8\%&73.9\%&16.3\%&14.8\%&97.5\%&66.8\%&20.9\%&23.0\%\cr
L$_{H\alpha}<10^{39}$&69.8\%&64.1\%&0.3\%&---&86.5\%&1.5\%&---&10.8\%&---&---\cr
L$_{H\alpha}>10^{39}$&87.2\%&5.9\%&62.8\%&---&19.7\%&90.6\%&---&9.4\%&---&---\cr
\tablenotetext{}{This sample presents the results of the contingency
table analysis. The percentages represent the probability that the
null hypothesis, that there is no correlation between the two quantities,
is right. Column 1: the way the sample was separated into subsamples;
Column 2: comparison between LINERs and Transition galaxies; Column 3:
LINERs and Seyferts, Column 4: LINERs and HII galaxies; Column 5: LINERs
and absorption line galaxies; Column 6: transition galaxies and Seyferts;
Column 7: transition galaxies and HII galaxies; Column 8: ransition galaxies
and absorption line galaxies; Column 9: Seyferts and HII galaxies;
Column 10: Seyferts and absorption line galaxies; Column 11: HII galaxies
and Absorption line galaxies.}
\enddata
\end{deluxetable}

\end{tiny}


\begin{references}

Antonucci, R. R. J. 1993, ARA\&A, 31, 473

Barnes, J.E., \& Hernquist, L., 1992. ARA\&A, 30, 70

Binney, J. \& Merrifield, M. 1998, in Galactic Astronomy, Princeton
University Press, p.164

Brown, M. J. I., Boyle, B. J. \& Webster, R. L. 2001, AJ, 121, 2381

Bushouse, H. A. 1986, AJ, 91, 255

Bushouse, H. A. 1987, ApJ, 320, 49

Byrd, G. G. \& Valtonen, M. 2001, AJ, 121, 2943

Byrd, G. G., Valtonen, M. J., Sundelius, B., Valtaoja, L. 1986, A\&A, 166, 75

Byrd, G. G., Sundelius, B., Valtonen, M. J. 1987, A\&A, 171, 16

Charlton, J. C. \& Salpeter, E. E. 1991, ApJ, 375, 517

Charlton, J. C., Whitmore, B. C. \& Gilmore, D. M., in Groups of Galaxies, ASP
Conference Series 70, Eds. O.-G. Richter, K. Borne, 1995, p. 49

Cid Fernandes, R. \& Terlevich, R. 1995, MNRAS, 272, 423

Cioni, M.-R. L., van der Marel, R. P., Loup, C. \& Habing, H. J. 2000, A\&A,
359, 601

Combes, F. 2001, astro-ph/0010570

Dahari, O. 1984, AJ, 89, 966

Dahari, O. 1985, AJ, 90, 1772

de Grijp, M. H. K., Keel, W. C., Miley, G. K., Goudfrooij, P. \& Lub, J.
1992, A\&AS, 96, 389

De Robertis, M. M., Yee, H. K. C. \& Hayhoe, K. 1998, ApJ, 496, 93

Dressler, A. 1980, ApJ, 235, 351

Dultzin-Hacyan, D., Krongold, Y., Fuentes-Guridi, I. \& Marziani, P. 1999,
ApJ, 513, L111

Falco, E. E. et al. 1999, PASP, 111, 438

Ferrarese, L. \& Merritt, D. 2000, ApJ, 539, L9

Fuentes-Willians, T. \& Stocke, J. T.  1988, AJ, 96, 1235

Gallimore, J. F. \& Keel, W. C. 1993,  AJ, 106, 1337

Garcia, A. M. 1993, A\&AS, 100, 47

Gebhardt, K. et al. 2000, ApJ, 543, L5

Gunn, J. 1979, in Active Galactic Nuclei, edited by C. Hazard \& S. Mitton,
(Cambridge University Press, Cambridge), p.213

Heckman, T. M. 1990, in Paired and Interacting Galaxies, IAU Colloquium
No. 124, eds. J. W. Sulentic \& W. C. Keel, NASA Conference Publication
3098, p.359

Heckman, T. M., Bothun, G. D., Balick, B. \& Smith, E. P. 1984, AJ, 89, 958

Heckman, T. M., Carty, T. J. \& Bothun, G. D. 1985, ApJ, 288, 122

Hernquist, L. 1989, Nature, 340, 687

Hernquist, L. \& Mihos, J. C. 1995, ApJ, 448, 41

Ho, L. C. 1999, ApJ, 516, 672

Ho, L. 1999, in ``Observational evidence for black holes in the Universe'',
ed. S. K. Chakrabarti (Dordrecht: Kluwer)

Ho, L. C., Filippenko, A. V. \& Sargent, W. L. W. 1997a, ApJS, 112, 315

Ho, L. C., Filippenko, A. V. \& Sargent, W. L. W. 1997b, ApJ, 487, 568

Ho, L. C., Filippenko, A. V. \& Sargent, W. L. W. 1997c, ApJ, 487, 591 

Ho, L. C. \& Peng, C. Y. 2001, ApJ, 555, 650

Ho, L. C. \& Ulvestad, J. S. 2001, ApJS, 133, 77

Huchra, J., Davis, M., Latham, D., Tonry, J. 1983, ApJS, 52, 89

Hutchings, J. 1983, PASP, 95, 799 

Impey, C. D., Bothum, G. D. \& Malin, D. 1988, ApJ, 330, 634

Impey, C. D., Sprayberry, D., Irwin, M. J. \& Bothum, G. D. 1996, ApJS, 105, 209

Karachentsev, I. 1987, Double Galaxies (Nauka, Moscow)

Keel, W. C., Kennicutt, R. C., Hummel, E. \& van der Hulst, J. M.
1985, AJ, 90, 708

Kennicutt, R. C., Keel, W. C., van der Hulst, J. M., Hummel, E. \& Roettiger,
K. A. 1987, AJ, 93, 1011

Krongold, Y., Dultzin-Hacyan, D., Marziani, P. 2001, AJ, 121, 702

Laurikainen, E. \& Salo, H. 1995, A\&A, 293, 683
 
Laurikainenn, E., Salo, H., Teerikorpi, P. \& Petrov, G. 1994, A\&AS, 108, 491

Lawrence, A., Rowan-Robinson, M., Leech, K., Jones, D. H. P. \& Wall, J. V.
1989, MNRAS, 240, 329

Lin, D. N. C., Pringle, J. E., Rees, M. J. 1988, ApJ, 328, 103

MacKenty, J. W. 1989, ApJ, 343, 125

Magorrian, J. et al. 1998, AJ, 115, 2285

Mateo, M. 1992. in The Stellar Populations of Galaxies: Proceedings
of the 149th IAU Symposium, eds. B. Barbuy and A. Renzini, Kluwer
Academic Publishers, Dordrecht, 1992, p.147

Mateo, M. L. 1998, ARA\&A, 36, 435

Mihos, J. C. \& Hernquist, L. 1994, ApJ, 425, L13

Noguchi, M. 1987, MNRAS, 228, 635

Noguchi, M. 1988, A\&A, 203, 259

Norman, C. 1987, in Galactic and Extragalactic star Formation, ed. R. E.
Pudritz \& M. Fich (Dordrecht: Kluwer), 495

Osterbrock, D. E. 1993, ApJ, 404, 551

Pastoriza, M. G., Donzelli, C. J. \& Bonatto, C. 1999, A\&A, 347, 55

Postman, M. \& Geller, M. J. 1984, ApJ, 281, 95

Press, W. H., Flannery, B. P., Teukolsky, S. A. \& Vettering, W. T.
1993, in Numerical Recipes in FORTRAN. The art of Scientific Computing
(Cambridge:University Press) p.476

Rafanelli, P., Violato, M. \& Baruffolo, A. 1995, AJ, 109, 1546

Sanders, D. B., Soifer, B. T., Elias, J. H., Madore, B. F., Matthews, K.,
Neugebauer, G. \& Scoville, N. Z. 1988, ApJ, 325, 74

Schmitt, H. R., Antonucci, R. R. J., Ulvestad, J. S., Kinney, A. L.,
Clarke, C. J. \& Pringle, J. E. 2001, ApJ, 555, 663

Schreier, E. J. et al. 2001, ApJ, in press

Schwartz, M. 1981, ApJ, 247, 77
 
Shlosman, I., Begelman, M. C. \& Frank, J. 1990, Nature, 345, 679

Stauffer, J. R. 1982, ApJ, 262, 66

Taniguchi, Y. \& Wada, K. 1996, ApJ, 469, 581

Taylor, C. L., Brinks, E., Grashuis, R. M. \& Skillman, E. D. 1995, ApJS, 99,
427

Telles, E. \& Maddox, S. 2000, MNRAS, 311, 307

Tully, R. B. 1988, Nearby Galaxies Catalog (Cambridge: Cambridge University
Press)

Ulvestad, J. S. \& Wilson, A. S. 1984, ApJ, 278, 544

Ulvestad, J. S. \& Wilson, A. S. 1985, ApJ, 285, 439

Whitmore, B. C., Gilmore, D. M. \& Jones, C. 1993, ApJ, 407, 489

Wright, G. S., Joseph, R. D., Robertson, N. A., James, P. A. \& Meikle,
W. P. S. 1988, MNRAS, 233, 1

\end{references}
\end{document}